\documentclass[aps,prl,twocolumn,groupedaddress,nofootinbib,longbibliography]{revtex4-2}
\usepackage{amsmath, amsfonts, amssymb,amsthm}
\usepackage{graphicx,subfigure}
\usepackage{mathrsfs}
\usepackage[mathscr]{eucal}
\usepackage[T1]{fontenc}
\usepackage{physics}
\usepackage{bm}
\usepackage{subfigure}
\usepackage[bookmarks=false]{hyperref}
\usepackage{xcolor}
\hypersetup{colorlinks=true,citecolor=blue,
linkcolor=blue,urlcolor=blue,pdfstartview=FitH,
bookmarksopen=true}

\newtheorem{theorem}{Theorem}

\begin{document}

\title{Inferring physical properties of symmetric states from the fewest copies}
\author{Da-Jian Zhang}
\email{zdj@sdu.edu.cn}
\affiliation{Department of Physics, Shandong University, Jinan 250100, China}
\author{D.~M.~Tong}
\email{tdm@sdu.edu.cn}
\affiliation{Department of Physics, Shandong University, Jinan 250100, China}

\date{\today}

\begin{abstract}
    Learning physical properties of high-dimensional states is crucial for developing quantum technologies but usually consumes an exceedingly large number of samples which are difficult to afford in practice. In this Letter, we use the methodology of quantum metrology to tackle this difficulty, proposing a strategy built upon entangled measurements for dramatically reducing sample complexity. The strategy, whose characteristic feature is symmetrization of observables, is powered by the exploration of symmetric structures of states which are ubiquitous in physics. It is provably optimal under some natural assumption, efficiently implementable in a variety of contexts, and capable of being incorporated into existing methods as a basic building block. We apply the strategy to different scenarios motivated by experiments, demonstrating exponential reductions in sample complexity.
\end{abstract}

\maketitle

Learning properties of states like expectation values of observables is of increasing importance in quantum physics, underpinning vast applications in emerging fields, such as quantum machine learning \cite{Biamonte2017}, quantum computational chemistry \cite{McArdle2020}, and variational quantum computing \cite{Cerezo2021,Tilly2022}. A fundamental difficulty in this task is that
any full reconstruction of a generic unknown state inevitably
consumes exponentially many samples due to the growth of the state space dimension with the system size. This makes traditional learning methods like full state tomography hopelessly inefficient and poses a serious challenge in the Noisy Intermediate-Scale Quantum (NISQ) era \cite{Preskill2018}.

Fortunately, most of the states that are relevant in quantum physics admit some nontrivial structures, such as those of low rank \cite{Gross2010,Liu2012}, matrix product states \cite{Cramer2010,Baumgratz2013}, and those with quasi-local structures \cite{Anshu2021,Rouze2021}.
Furthermore, for many purposes, one is only interested in some specific properties of states, e.g., mean energies of many-body systems, which makes it unnecessary to fully reconstruct states for obtaining all their information. These well-motivated physical considerations have led to a series of proposals for more efficient alternatives to full state tomography, with excellent examples including compressed sensing \cite{Gross2010,Liu2012}, adaptive tomography \cite{Mahler2013,Qi2017}, self-guided tomography \cite{Ferrie2014,Rambach2021}, and classical shadows  \cite{Aaronson2018,Huang2020}.

While existing methods generally rely on local measurements, it has been recognized that entangled measurements are typically far more efficient than local measurements for extracting information from unknown states \cite{Peres1991,Gisin1999,Huang2022,Aharonov2022,MiguelRamiro2022,Larocca2022}. Moreover, the ongoing development of large-scale quantum computers opens up exciting possibilities of leveraging quantum computational resources to realize entangled measurements \cite{Huang2022,Aharonov2022}. In particular, motivated by the availability of NISQ computers \cite{Bharti2022}, a number of experiments \cite{Tang2020,Huang2022a,Conlon2023} have been carried out for realizing entangled measurements as well as demonstrating their superiority over local measurements. An important issue following is therefore to explore the usefulness of entangled measurements in learning properties of states \cite{Huang2022,Aharonov2022}.

Here we propose a strategy built upon entangled measurements, for dramatically reducing sample complexity beyond what can be achieved with local measurements. The basic idea is to explore symmetric structures of states which are ubiquitous in quantum physics and known \textit{a priori} in various situations. Using information-theoretic tools from quantum metrology \cite{1976Helstrom,2011Holevo}, we figure out the measurement that can make optimal use of these symmetric structures for learning expectation values of observables. This enables our strategy to operate at the optimal sample efficiency in a variety of physical contexts, thereby achieving a sought-after goal in quantum metrology \cite{1994Braunstein3439,2020Zhang23418,Zhang2022}, which is unlikely, if not impossible, to reach with local measurements.
Further, we apply our strategy to two physical contexts involving translational and permutational symmetries, showing that it allows for saving exponentially many samples while merely consuming polynomial amounts of quantum computational resources. The findings of this Letter uncover an intriguing route to reducing sample complexity via taking advantage of symmetric structures of states, which opens opportunities for leveraging recent breakthroughs on large-scale quantum computers to facilitate a plethora of learning tasks.

We first recall some basic facts. A common task in quantum physics is to measure the expectation value $\overline{X}=\tr(X\rho)$ of an observable $X$ in a state $\rho$.
Traditionally, $\overline{X}$ is obtained from the projective measurement of $X$, provided that this measurement is not difficult to implement in experiments. As the measurement outcomes are the eigenvalues of $X$, there is some uncertainty in obtaining $\overline{X}$, which can be characterized by the variance $(\Delta X)^2=\overline{(X-\overline{X})^2}$. More precisely, the uncertainty is $(\Delta X)^2/M$ if the measurement is repeated $M$ times. So, to measure $\overline{X}$ up to a certain prescribed tolerance $\epsilon$, the number of measurements needed and hence samples consumed is
\begin{eqnarray}\label{M-X}
    M_X=\left\lceil\frac{(\Delta X)^2}{\epsilon}\right\rceil,
\end{eqnarray}
where $\lceil \cdot \rceil$ denotes the ceiling function.

We now consider the setting that we know \textit{a priori} the symmetric structures of $\rho$, which can be described by a finite or compact Lie group $G$, that is,  $U_g\rho U_g^\dagger=\rho$ for $g\in G$, where $U_g$ denotes a unitary representation of ${G}$. It is worth noting that the above consideration covers the vast majority of symmetries of interest in quantum physics.
Then a question of relevance in a startling variety of contexts is: How can we exploit the symmetric structures of $\rho$ to reduce \textit{to the greatest extent} the number of samples required in measuring the expectation value $\overline{X}$ of \textit{any given observable} $X$ in $\rho$ up to the prescribed tolerance $\epsilon$?

In this Letter, we will answer the above question by showing that
the projective measurement of another observable
\begin{eqnarray}\label{opt-Y}
    Y=\mathcal{T}(X)
\end{eqnarray}
can not only be alternatively employed to measure $\overline{X}$ but also maximally reduce the number of samples required by taking advantage of the symmetric structures of $\rho$.
Here, $\mathcal{T}$ is the so-called $G$-twirling operation defined as
\begin{eqnarray}\label{TO}
    \mathcal{T}(X)={\abs{G}}^{-1}\sum_{g\in G}U_g X U_g^\dagger,
\end{eqnarray}
for a finite group $G$, with $\abs{G}$ being the number of elements in $G$. When $G$ is a compact Lie group,
$\mathcal{T}(X)=\int_{G}\mathrm{d}\nu(g) U_g X U_g^\dagger$, where $\nu(g)$ denotes the normalized Haar measure \cite{1950Halmos}.
It is interesting to note that $Y$ satisfies $[Y,U_g]=0$ for $g\in G$ and is therefore the symmetrized counterpart of $X$.
We emphasize that $Y$ is always different from $X$ except when $X$ commutes with all $U_g$.

We first show that $X$ and $Y$ have the following subtle relations.
\begin{theorem}\label{Theorem1}
    The observable $Y$ shares the same expected value with $X$ but with a generally smaller variance, i.e., $\overline{Y}=\overline{X}$ but $(\Delta Y)^2\leq(\Delta X)^2$.
\end{theorem}
Physically, the equality $\overline{Y}=\overline{X}$ means that the projective measurement of $Y$ can be an alternative to the projective measurement of $X$ for obtaining $\overline{X}$. The inequality $(\Delta Y)^2\leq(\Delta X)^2$ implies that the former generally consumes fewer samples than the latter for reaching the same measurement precision, that is, $M_Y\leq M_X$, where
\begin{eqnarray}\label{M-Y}
    M_Y=\left\lceil\frac{(\Delta Y)^2}{\epsilon}\right\rceil
\end{eqnarray}
is the number of samples required in the projective measurement of $Y$.

We now prove Theorem \ref{Theorem1}. Let us consider the case that $G$ is a finite group. Using the cyclic property of the trace and noting that $[\rho, U_g]=0$ for $g\in G$, we have $\overline{U_gXU_g^\dagger}=\overline{X}$. In conjunction with Eq.~(\ref{TO}), this further leads to $\overline{Y}=\sum_g\overline{U_gXU_g^\dagger}/\abs{G}=\overline{X}$. To prove $(\Delta Y)^2\leq(\Delta X)^2$, we make use of the known result that, for two observables $A$ and $B$, there is $\Delta(A+B)\leq\Delta A+\Delta B$ \cite{Pati2007,Maccone2014}. Using this result and noting that $\Delta(U_gXU_g^\dagger)=\Delta X$, we have that $\Delta Y=\Delta(\abs{G}^{-1}\sum_gU_gXU_g^\dagger)\leq\abs{G}^{-1}\sum_g\Delta(U_gXU_g^\dagger)=\Delta X$. Then the proof of Theorem \ref{Theorem1} is completed by further noting that the above reasoning can be carried over straightforwardly to the case of $G$ being a compact Lie group.

We then address the optimality of the projective measurement of $Y$. Note that there are many other measurement strategies that can be used to measure $\overline{X}$, e.g., classical shadows based on randomized measurements \cite{Aaronson2018,Huang2020}.
Without loss of generality, any strategy for measuring $\overline{X}$ can be described as first performing a measurement on $\rho^{\otimes M}$ and then classically post-processing the measurement outcome to yield an estimate of $\overline{X}$ \cite{1976Helstrom,2011Holevo}. Here $M$ denotes the number of samples consumed in the strategy in question. Any measurement can be described by a positive operator-valued measure (POVM) $\{\Pi_{\bm{x}}\}$ satisfying $\sum_{\bm{x}} \Pi_{\bm{x}}=\openone$, where $\bm{x}$ labels the measurement outcome and could be multivariate in general. Any classical post-processing of $\bm{x}$ amounts to a map $\overline{X}_\textrm{est}(\bm{x})$, which takes $\bm{x}$ as the input and outputs an estimate of $\overline{X}$. The uncertainty in measuring $\overline{X}$ can be quantified by the variance $\sum_{\bm{x}}\tr(\Pi_{\bm{x}}\rho^{\otimes M})\left[\overline{X}_\textrm{est}(\bm{x})-\overline{X}\right]^2$ \cite{1976Helstrom,2011Holevo}. The requirement of measuring $\overline{X}$ up to $\epsilon$ amounts to demanding
\begin{eqnarray}\label{req-M}
    \epsilon\geq\sum_{\bm{x}}\tr(\Pi_{\bm{x}}\rho^{\otimes M})\left[\overline{X}_\textrm{est}(\bm{x})-\overline{X}\right]^2,
\end{eqnarray}
which imposes constraints on $M$ so that $M$ cannot be arbitrarily small. We uncover a fundamental bound on $M$ as follows.
\begin{theorem}\label{Theorem2}
    The number $M$ of samples consumed in any strategy for measuring $\overline{X}$ up to $\epsilon$ is bounded as $M\geq M_Y$ if nothing about $\rho$ is known except its symmetric structures.
\end{theorem}

Theorem \ref{Theorem2} means that the projective measurement of $Y$ consumes the fewest copies of $\rho$ for measuring $\overline{X}$ up to $\epsilon$, under the assumption that our knowledge about $\rho$ is only its symmetric structures. That is, the projective measurement of $Y$ has the optimal sample efficiency allowed by quantum mechanics whenever the assumption holds.

Our proof of Theorem \ref{Theorem2} is based on the quantum Cram\'{e}r-Rao bound \cite{1976Helstrom,2011Holevo} (see Supplemental Material \cite{SM} for details). To calculate this bound, the key step in our proof is to write out a general expression of $\rho$ based on the known symmetric structures. According to the representation theory of groups \cite{1962Curtis}, $U_g$ can be expressed as
\begin{eqnarray}\label{Tong1}
    U_g\cong\bigoplus_{\alpha=1}^s \openone_{n_\alpha}\otimes U_\alpha(g),
\end{eqnarray}
where $U_\alpha(g)$ is the $\alpha$-th irreducible representation of $G$ with dimension $d_\alpha$ and multiplicity $n_\alpha$, and $\openone_{n_\alpha}$ is the $n_\alpha\times n_\alpha$ identity matrix. As $[\rho,U_g]=0$ for $g\in G$, $\rho$ must be of the form
\begin{eqnarray}\label{decom-rho-main}
    \rho\cong\bigoplus_{\alpha=1}^s q_\alpha\rho_\alpha\otimes\frac{\openone_{d_\alpha}}{d_\alpha},
\end{eqnarray}
where $q_\alpha\geq 0$ satisfies $\sum_{\alpha=1}^s q_\alpha=1$ and $\rho_\alpha$ denotes a density matrix. Using the generalized Bloch representation \cite{2003Kimura339}, we can further express $\rho_\alpha$ as (see Sec.~II of Ref.~\cite{SM})
\begin{eqnarray}\label{alpha-rho-main}
    \rho_\alpha=\frac{\openone_{n_\alpha}}{n_\alpha}+\frac{1}{2}\bm{r}_\alpha\cdot\bm{\lambda}_\alpha,
\end{eqnarray}
where $\bm{r}_\alpha$ and $\bm{\lambda_\alpha}$ are used to denote the generalized Bloch vector and the generators of Lie algebra $\mathfrak{su}(n_\alpha)$, respectively \cite{SM}.
It follows from Eqs.~(\ref{decom-rho-main}) and (\ref{alpha-rho-main}) that $\rho$ is explicitly expressed in terms of the parameters $\{q_\alpha,\bm{r}_\alpha\}_{\alpha=1}^s$.
This allows us to figure out the quantum Cram\'{e}r-Rao bound,
\begin{eqnarray}\label{QCRB-main}
    \sum_{\bm{x}}\tr(\Pi_{\bm{x}}\rho^{\otimes M})\left[\overline{X}_\textrm{est}(\bm{x})-\overline{X}\right]^2\geq\frac{1}{MJ[\overline{X};\rho]},
\end{eqnarray}
where $J[\overline{X};\rho]$ denotes the quantum Fisher information about $\overline{X}$ given $\rho$ (see Sec.~V of Ref.~\cite{SM} for its explicit expression). The meaning of Eq.~(\ref{QCRB-main}) is that the uncertainty in any strategy for measuring $\overline{X}$ is fundamentally constrained by $\frac{1}{MJ[\overline{X};\rho]}$, irrespective of the choices of $\{\Pi_{\bm{x}}\}$ and $\overline{X}_\textrm{est}(\bm{x})$ \cite{1976Helstrom,2011Holevo}.
Using Eqs.~(\ref{req-M}) and (\ref{QCRB-main}), we obtain
\begin{eqnarray}\label{bound-M-main}
    M\geq\frac{1}{\epsilon J[\overline{X};\rho]}.
\end{eqnarray}
Besides, we derive, after some algebra (see Sec.~VI of Ref.~\cite{SM}),
\begin{eqnarray}\label{Y-QFI-main}
    (\Delta Y)^2=\frac{1}{J[\overline{X};\rho]}.
\end{eqnarray}
The proof of Theorem \ref{Theorem2} is completed by inserting Eq.~(\ref{Y-QFI-main}) into Eq.~(\ref{bound-M-main}).

\begin{figure}
    \includegraphics[width=0.47\textwidth]{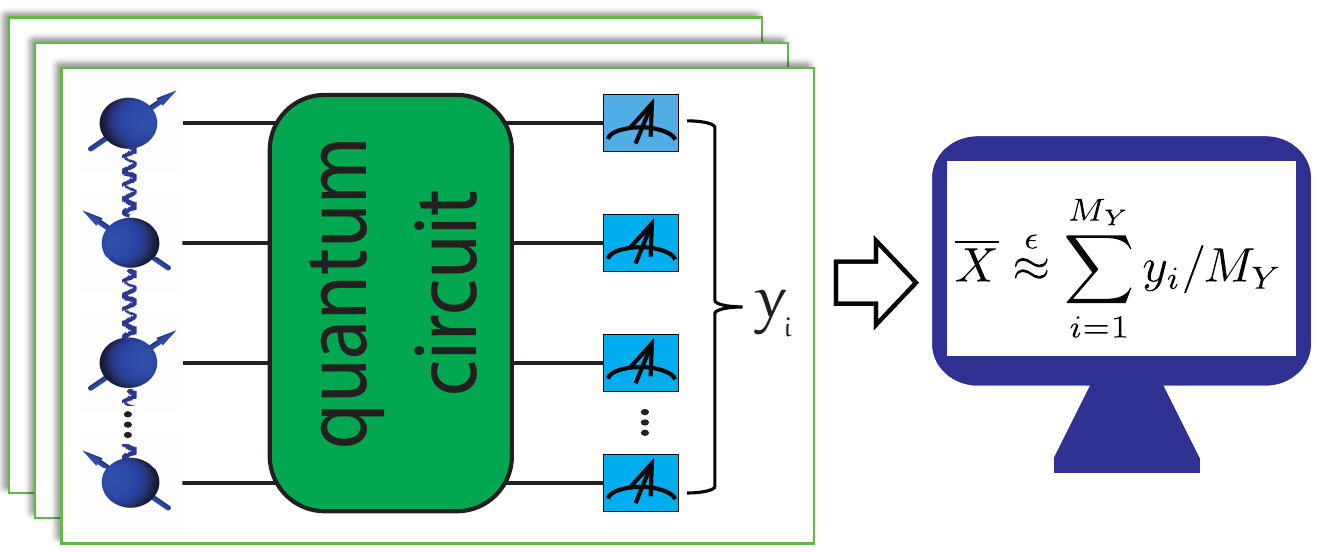}
    \caption{Schematic of our strategy. To measure the expectation value $\overline{X}$ of an observable $X$ in $\rho$ which is possibly a many-body state, we perform the projective measurement of the observable $Y$ on each sample. This measurement can be implemented by utilizing the quantum circuit $V$ to transform the eigenbasis of $Y$ into the computational basis and then performing the standard measurement in the computational basis. Repeating this procedure $M_Y=\lceil(\Delta Y)^2/\epsilon\rceil$ times, we can obtain $\overline{X}$ up to a prescribed tolerance $\epsilon$ by averaging the outcomes $y_i$.}
    \label{fig1}
\end{figure}

With the above analysis, we are ready to specify our strategy. \textit{To measure $\overline{X}$, our strategy is to repeatedly perform the projective measurement of $Y$ on $M_Y$ samples.} This allows for yielding $\overline{X}$ up to $\epsilon$ after averaging the $M_Y$ outcomes, i.e., $\overline{X}\overset{\epsilon}{\approx}\sum_{i=1}^{M_Y}y_i/M_Y$, where $y_i$ denotes the $i$-th measurement outcome and is an eigenvalue of $Y$. Interestingly, as demonstrated below, the projective measurement of $Y$ is often an entangled measurement \cite{Tang2020,Huang2022a,Conlon2023}, since the eigenbasis of $Y$ may contain entangled states. To implement this measurement, we can first apply a quantum circuit $V$ to transform the eigenbasis of $Y$ into the computational basis and then perform the standard measurement in the computational basis (see Fig.~\ref{fig1}). So, the working principle of our strategy is to leverage quantum computational resources to reduce sample complexity.
Notably, our strategy operates at the optimal sample efficiency allowed by quantum mechanics and outperforms other measurement strategies in terms of sample complexity whenever the assumption in Theorem \ref{Theorem2} holds.

{\textit{Application  1: translational symmetries.}} Let us apply our strategy to the scenario that $\rho$ is a state of $n$ qubits with the symmetric structures described by the translation group $G=\{T^i, i=0,\cdots,2^n-1\}$. Here, $T$ is defined as $T\ket{j}=\ket{j+1}$ with the periodic boundary condition $\ket{2^n}=\ket{0}$, where $\ket{j}$, $j=0,\cdots,2^n-1$, denote the computational basis. Note that translational symmetries are ubiquitous in condensed-matter physics. The above scenario could arise, e.g., in quantum simulations of electrons in crystalline solids \cite{Bloch2012,Georgescu2014}, for which Bloch's theorem states that the solutions to the Schr\"{o}dinger equations are Bloch states and hence respect translational symmetries.

To show the usefulness of our strategy, we calculate $(\Delta X)^2/(\Delta Y)^2$, which characterizes the ratio between $M_X$ and $M_Y$. Using $[\rho,T]=0$ and noting that the eigenbasis of $T$ is the Fourier basis $\ket{f_j}=\sum_ke^{2\pi \mathrm{i} jk/2^n}\ket{k}/{\sqrt{2^{n}}}$, we can write $\rho$ as $\rho=\sum_jp_j\ket{f_j}\bra{f_j}$, where $p_j\geq 0$ satisfies $\sum_jp_j=1$. Then, expressing $X$ as
$X=\sum_{jk}X_{jk}\ket{f_j}\bra{f_k}$
with $X_{jk}=\bra{f_j}X\ket{f_k}$, we have
\begin{eqnarray}
    (\Delta X)^2=(\sum_{jk}p_j\abs{X_{jk}}^2)-(\sum_jp_jX_{jj})^2.
\end{eqnarray}
Besides, using Eq.~(\ref{TO}) and $T\ket{f_j}=e^{-2\pi \mathrm{i}j/2^n}\ket{f_j}$, we have
$Y=\sum_j X_{jj}\ket{f_j}\bra{f_j}$,
which further leads to
\begin{eqnarray}
    (\Delta Y)^2=(\sum_jp_j{X}_{jj}^2)-(\sum_jp_jX_{jj})^2.
\end{eqnarray}
Hence,
$(\Delta X)^2-(\Delta Y)^2=\sum_{j}\sum_{k\neq j}p_j\abs{X_{jk}}^2$
contains exponentially many nonnegative terms. Roughly speaking, this indicates that $(\Delta X)^2/(\Delta Y)^2$ is very large for countless choices of $X$.
To clearly see this point, we take
\begin{eqnarray}
    X=\bigotimes_{l=1}^n\frac{1}{\sqrt{2}}(\sigma_x^l+\sigma_z^l)
\end{eqnarray}
as an example,
where $\sigma_\alpha^l$, $\alpha=x,y,z$, are the Pauli matrices acting on the $l$-th qubit. Direct calculations show \cite{SM} that
\begin{eqnarray}
    (\Delta X)^2/(\Delta Y)^2\geq 2^n-1,
\end{eqnarray}
implying that the reduction allowed by our strategy can be exponential in $n$.

We point out that our strategy is efficiently implementable on a quantum computer in the scenario under consideration. Indeed, the eigenbasis of $Y$ is the Fourier basis, which implies that the quantum circuit $V$ is just the inverse quantum Fourier transform. That our strategy is efficiently implementable follows from the known result that the inverse quantum Fourier transform can be realized as a quantum circuit consisting of only $\mathcal{O}(n^2)$ Hadamard gates and controlled phase shift gates \cite{2010Nielsen} (see also Ref.~\cite{Griffiths1996} for a semiclassical realization without using two-bit gates).

\textit{Application  2: permutational symmetries.} Let us consider again $n$ qubits but in the state whose symmetric structures are described by the permutation group $G=\{P_s, s\in S_n\}$. Here, $s$ labels a permutation in the symmetric group $S_n$ and $P_s$ is defined by $P_s\ket{\psi_1}\otimes\cdots\otimes\ket{\psi_n}=\ket{\psi_{s(1)}}\otimes\cdots\otimes\ket{\psi_{s(n)}}$. This scenario arises frequently in multipartite experiments \cite{2007Kiesel63604,2009Wieczorek20504,2009Prevedel20503}, in which the states involved are typically invariant under permutations  \cite{2010Toth250403}. For example, three well-known states of this type are Werner states \cite{1989Werner4277}, Dicke states \cite{Dicke1954}, and Greenberger–Horne–Zeilinger (GHZ) states \cite{Greenberger1989}, which are key resources in quantum information processing  \cite{2009Horodecki865,2009Guehne1,2017Streltsov41003,2018Hu1}. Below, motivated by the fact that Pauli measurements are widely used in multipartite experiments, we explore our strategy to reduce sample complexity in Pauli measurements.

We can express a generic Pauli observable as
\begin{eqnarray}
    X_{\bm{kl}}=\sigma_x^{k_1}\sigma_z^{l_1}\otimes\cdots\otimes\sigma_x^{k_n}\sigma_z^{l_n}(\mathrm{i})^{\bm{k}\cdot\bm{l}},
\end{eqnarray}
where $\bm{k}=(k_1,\cdots,k_n)$ and $\bm{l}=(l_1,\cdots,l_n)$ are two vectors of binary numbers and $\bm{k}\cdot\bm{l}=\sum_{i=1}^nk_il_i$ denotes the usual dot product.
Note that, associated to each $X_{\bm{kl}}$, there is a symmetrized counterpart $Y_{\bm{kl}}=\mathcal{T}(X_{\bm{kl}})$. To illustrate the superiority of the projective measurement of $Y_{\bm{kl}}$ over the Pauli measurement of $X_{\bm{kl}}$, we evaluate $(\Delta X_{\bm{kl}})^2$ and $(\Delta Y_{\bm{kl}})^2$ on the GHZ state of $n$ qubits. Hereafter we assume for simplicity that $n$ is odd. It can be shown that
\begin{eqnarray}
    (\Delta X_{\bm{kl}})^2=1
\end{eqnarray}
for any Pauli observable with $\abs{\bm{k}}\neq 0$ and $n$ \cite{SM}. Here $\abs{\bm{k}}=\sum_{i=1}^nk_i$. By contrast,
\begin{eqnarray}
    (\Delta Y_{\bm{kl}})^2=1\Big/\binom{n}{\abs{\bm{k}}}
\end{eqnarray}
for the same Pauli observable \cite{SM}. To clearly see the difference between $(\Delta X_{\bm{kl}})^2$ and $(\Delta Y_{\bm{kl}})^2$, we consider the Pauli measurements with
\begin{eqnarray}
    (1-\delta)\frac{n}{2}<\abs{\bm{k}}<(1+\delta)\frac{n}{2},
\end{eqnarray}
referred to as the typical Pauli measurements. Here, $0<\delta<1$ is fixed. We show that
\begin{eqnarray}
    \frac{(\Delta X)^2}{(\Delta Y)^2}
    \geq\sqrt{\frac{2}{n\pi(1-\delta^2)}}\left[\frac{4}{(1-\delta)^{1-
                    \delta}(1+\delta)^{1+
                    \delta}}\right]^{\frac{n}{2}}
\end{eqnarray}
for any typical Pauli measurement \cite{SM}. Noting that $(1-\delta)^{1-\delta}(1+\delta)^{1+\delta}< 4$ for $0<\delta<1$, we deduce that $(\Delta X)^2/(\Delta Y)^2$ is exponential in $n$. Besides, the number of typical Pauli measurements is $\geq 4^n(1-\frac{1}{n\delta^2})$ \cite{SM}, which means most of Pauli measurements are typical for a large $n$. So, our strategy allows for exponentially reducing sample complexity for most of Pauli measurements when $n$ is large.

Notably, our strategy can be efficiently implemented on a quantum computer in the considered scenario, too. Indeed, as detailed in Ref.~\cite{SM}, the eigenbasis of $Y$ can be mapped into the computational basis via the quantum Schur transform followed by at most $\lceil \frac{n}{2} \rceil$ controlled gates. That our strategy is efficiently implementable follows from the known result that the quantum Schur transform can be realized as a quantum circuit of polynomial size \cite{Bacon2006}.

%\textcolor{red}{{\textit{Concluding remarks.}}}---
Before concluding, we remark that while the assumption in Theorem \ref{Theorem2} holds in numerous scenarios, there are also many cases in which we know other structures of $\rho$ besides its symmetric structures. For example, apart from translational symmetries, a many-body state usually admits quasi-local structures, based on which some learning methods have been proposed \cite{Anshu2021,Rouze2021}. As such, our strategy could be incorporated into existing methods as a basic building block for further reducing sample complexity by exploring symmetric structures of states.
Besides, it should be noted that there are difficulties in implementing the projective measurement of $X$ in many situations. For example, it was recognized that the mean energy of a many-body system is considerably difficult to measure, due to which only a few architecture-specific experiments have been carried out so far \cite{Safranek2023}. Interestingly, our strategy allows for inferring $\overline{X}$ from the projective measurement of $Y$ which, as demonstrated above, may be efficiently implementable even when this is not the case for $X$.

In conclusion, targeting at measuring expectation values of observables,  we have proposed an entangled-measurement-based strategy that can leverage quantum computational resources to dramatically reduce sample complexity. Our strategy, which is powered by the exploration of symmetric structures of states, is to infer the expectation value $\overline{X}$ of an observable $X$ from the projective measurement of its symmetrized counterpart $Y$ in Eq.~(\ref{opt-Y}) rather than $X$ itself. This allows our strategy to be an intriguing alternative with the optimal sample efficiency in a variety of contexts, as stated in Theorems \ref{Theorem1} and \ref{Theorem2}.

To illustrate the significance of our strategy, we have applied it to two scenarios involving different kinds of symmetric structures of states, i.e., those described respectively by the translation and permutation groups, which are ubiquitous in condensed-matter physics and quantum many-body physics. These applications clearly illustrate that our strategy allows for yielding exponential reductions in sample complexity while merely consuming polynomial amounts of quantum computational resources.

The present Letter opens many interesting topics for future work, e.g., how to optimally take advantage of symmetric structures of states to reduce the sample complexity in simultaneously measuring expectation values of multiple observables \cite{2018Zhang170501,Huggins2022} and further to extend the scope of discussions from expectation values of observables to other properties of states like various resource measures \cite{2009Horodecki865,2009Guehne1,2017Streltsov41003,2018Hu1}.

\begin{acknowledgments}
    This work was supported by the National Natural Science Foundation of
    China through Grant Nos.~12275155 and 12174224.
\end{acknowledgments}

\renewcommand{\theequation}{\thesubsection S\arabic{equation}}
\setcounter{equation}{0}

\def\eqB{{1}}
\def\eqC{{3}}
\def\eqY{{4}}

\def\eqqfi{{10}}

\begin{center}
    \huge{\textbf{Supplemental Material}}
\end{center}

\section{I. Fundamentals of quantum metrology}\label{sec1}

Here we recall some fundamentals of quantum metrology, whose mathematical foundation is the quantum parameter estimation theory \cite{1976Helstrom,2011Holevo}. The theory deals with the quantum parameter estimation problems, i.e., the problems in which both the state in question and the quantity to be estimated are characterized by some unknown parameters.
Specifically, the state in question can be a density matrix $\rho(\bm{\theta})$ characterized by $\mathfrak{p}$ unknown parameters
$\bm{\theta}=(\theta_1,\cdots,\theta_\mathfrak{p})$,
and the quantity to be estimated could be a function of $\bm{\theta}$, denoted as $\beta(\bm{\theta})$. Given $M$ copies of $\rho(\bm{\theta})$, the parameter estimation problem is to infer the value of $\beta$ by performing a measurement on $\rho(\bm{\theta})^{\otimes M}$.

Any measurement can be described by a positive operator-valued measure (POVM) $\Pi_{\bm{x}}$ satisfying $\sum_x \Pi_{\bm{x}}=\openone$, where $\bm{x}$ labels the outcome of the measurement and could be multivariate in general.
Any inference rule amounts to an estimator
${\beta}_\textrm{est}({\bm{x}})$, which takes ${\bm{x}}$ as the input and outputs an estimate of $\beta$. It is a natural requirement for an estimator $\beta_\textrm{est}({\bm{x}})$ to be unbiased, that is,
\begin{eqnarray}
    \sum_{\bm{x}}p_{\bm{\theta}}({\bm{x}})\beta_\textrm{est}(x)=\beta,
\end{eqnarray}
where
\begin{eqnarray}
    p_{\bm{\theta}}({\bm{x}})=\tr[\Pi_{\bm{x}}\rho(\bm{\theta})^{\otimes M}]
\end{eqnarray}
denotes the probability of getting outcome ${\bm{x}}$. The estimation error can be quantified by the variance of the estimator $\beta_\textrm{est}$ \cite{1976Helstrom,2011Holevo},
\begin{eqnarray}\label{est-error}
    \textrm{Var}[\beta_\textrm{est}]=\sum_{\bm{x}}p_{\bm{\theta}}({\bm{x}})[\beta_\textrm{est}({\bm{x}})-\beta]^2.
\end{eqnarray}

One celebrated result in quantum metrology is the quantum Cram\'{e}r-Rao theorem \cite{1976Helstrom,2011Holevo,1994Braunstein3439}. It states that  $\textrm{Var}[\beta_\textrm{est}]$ is bounded as
\begin{eqnarray}\label{SM-QCRB}
    \textrm{Var}[\beta_\textrm{est}]\geq\frac{1}{MJ[\beta;\rho(\bm{\theta})]},
\end{eqnarray}
known as the quantum Cram\'{e}r-Rao bound (QCRB).
Here, $J[\beta;\rho(\bm{\theta})]$ is the quantum Fisher information (QFI) about $\beta$ when the state in question is $\rho(\bm{\theta})$. $J[\beta;\rho(\bm{\theta})]$ is defined as
\begin{eqnarray}
    J[\beta;\rho(\bm{\theta})]=\frac{1}{\partial\beta [H(\bm{\theta})]^{-1}\partial\beta^T},
\end{eqnarray}
where
\begin{eqnarray}
    \partial\beta=\left(\frac{\partial\beta}{\partial\theta_1},\cdots,\frac{\partial\beta}{\partial
        \theta_\mathfrak{p}}\right)
\end{eqnarray}
is a $\mathfrak{p}$-dimensional row vector
and $H(\bm{\theta})$ is a $\mathfrak{p}\times\mathfrak{p}$
symmetric matrix known as the QFI matrix. The $kl$ element of $H(\bm{\theta})$ is
\begin{eqnarray}
    H_{kl}=\tr[\rho(\bm{\theta})(L_k\circ L_l)],
\end{eqnarray}
where
\begin{eqnarray}
    L_k\circ L_l=(L_kL_l+L_lL_k)/2
\end{eqnarray}
denotes the Jordan product, and $L_k$ is the symmetric logarithmic derivative defined as the Hermitian operator satisfying
\begin{eqnarray}
    \frac{\partial}{\partial\theta_k}\rho(\bm{\theta})=L_k\circ\rho(\bm{\theta}).
\end{eqnarray}
The physical meaning of the QCRB is that quantum mechanics does not allow for estimating $\beta$ with arbitrarily high precision. Instead, the estimation error produced in any strategy for estimating $\beta$ with the $M$ copies of $\rho(\bm{\theta})$ is fundamentally constrained by the QCRB, which represents the ultimate precision allowed by quantum mechanics \cite{1994Braunstein3439}.

We remark that the QCRB also holds when $\beta_\textrm{est}(x)$ is biased, provided that the estimation error in Eq.~(\ref{est-error}) is redefined to remove the local difference
in the ``units'' of the estimate $\beta_\textrm{est}(x)$ and the value $\beta$ (see Ref.~\cite{1994Braunstein3439} for details).

\section{II. Explicit expression of $\rho$}\label{sec3}

Here we write out an explicit expression of the $\rho$ that satisfies $[\rho,U_g]=0$ for all $g\in G$.
According to the representation theory of groups \cite{1962Curtis}, $U_g$ can always be expressed as
\begin{eqnarray}\label{Tong1}
    U_g\cong\bigoplus_{\alpha=1}^s I_{n_\alpha}\otimes U_\alpha(g),
\end{eqnarray}
where $U_\alpha(g)$ denotes the $\alpha$-th irreducible representation of $G$ with dimension $d_\alpha$ and multiplicity $n_\alpha$. Here, $\cong$ means that these equalities are up to a unitary transformation. We deduce from $[\rho,U_g]=0$ that $\rho$ can be expressed as
\begin{eqnarray}\label{rho-exp1}
    \rho\cong\bigoplus_{\alpha=1}^s A_\alpha\otimes\openone_{d_\alpha},
\end{eqnarray}
where $A_\alpha$ is an $n_\alpha\times n_\alpha$ semipositive matrix. Noting that $A_\alpha/\tr A_\alpha$ is a density matrix and $\tr\rho=1$, we can rewrite Eq.~(\ref{rho-exp1}) as
\begin{eqnarray}\label{decom-rho}
    \rho\cong\bigoplus_{\alpha=1}^s q_\alpha\rho_\alpha\otimes\frac{\openone_{d_\alpha}}{d_\alpha},
\end{eqnarray}
where $q_\alpha\geq 0$ satisfies $\sum_{\alpha=1}^s q_\alpha=1$ and $\rho_\alpha$ denotes a density matrix.

Note that any traceless $n_\alpha\times n_\alpha$ matrix can be expressed in terms of the generators of Lie algebra $\mathfrak{su}(n_\alpha)$. We collectively denote the generators of Lie algebra $\mathfrak{su}(n_\alpha)$ as
\begin{eqnarray}
    \bm{\lambda}_\alpha=(\lambda_{\alpha,1},\lambda_{\alpha,2},\cdots,\lambda_{\alpha,n_\alpha^2-1}),
\end{eqnarray}
where $\lambda_{\alpha,i}$, $i=1,2,\cdots,n_\alpha^2-1$, are $n_\alpha\times n_\alpha$ matrices.
It has been shown \cite{2003Kimura339} that $\lambda_{\alpha,i}$ can be chosen to satisfy
\begin{eqnarray}\label{SM-p1}
    \lambda_{\alpha,i}^\dagger=\lambda_{\alpha,i},~~
    \tr\lambda_{\alpha,i}=0,~~
    \tr(\lambda_{\alpha,i}\lambda_{\alpha,j})=2\delta_{ij},
\end{eqnarray}
and be characterized by structure constants $f_{\alpha,ijk}$ and $g_{\alpha,ijk}$ as
\begin{eqnarray}\label{SM-p2}
    [\lambda_{\alpha,i},\lambda_{\alpha,j}]=2i\sum_kf_{\alpha,ijk}\lambda_{\alpha,k},
\end{eqnarray}
and
\begin{eqnarray}\label{SM-p3}
    \{\lambda_{\alpha,i},\lambda_{\alpha,j}\}=4\delta_{ij}\frac{\openone_{n_\alpha}}{n_\alpha}+2\sum_kg_{\alpha,ijk}\lambda_{\alpha,k},
\end{eqnarray}
where $\{\cdot,\cdot\}$ denotes the anticommutator. Note that $f_{\alpha,ijk}$ is completely antisymmetric and $g_{\alpha,ijk}$ is completely symmetric. The generators $\bm{\lambda}_\alpha$ satisfying Eqs.~(\ref{SM-p1}), (\ref{SM-p2}), and (\ref{SM-p3}) can be regarded as generalizations of the Pauli matrices.
We can then expand $\rho_\alpha$ in terms of the generators as
\begin{eqnarray}\label{alpha-rho}
    \rho_\alpha=\frac{\openone_{n_\alpha}}{n_\alpha}+\frac{1}{2}\bm{r}_\alpha\cdot\bm{\lambda}_\alpha.
\end{eqnarray}
Here
\begin{eqnarray}
    \bm{r}_\alpha=(r_{\alpha,1},r_{\alpha,2},\cdots,r_{\alpha,n_\alpha^2-1})
\end{eqnarray}
is a $(n_\alpha^2-1)$-dimensional real vector known as the generalized Bloch vector \cite{2003Kimura339} .

From Eqs.~(\ref{decom-rho}) and (\ref{alpha-rho}), it follows that $\rho$ is explicitly expressed in terms of the parameters
\begin{eqnarray}\label{unknown-para}
    \bm{\theta}=({\bm{q}},\bm{r}_1,\cdots,\bm{r}_s),
\end{eqnarray}
with
\begin{eqnarray}
    {\bm{q}}=(q_1,\cdots,q_{s-1}),
\end{eqnarray}
where in view of the constraint that $\sum_{\alpha=1}^sq_\alpha=1$, we have chosen $q_1,\cdots,q_{s-1}$ as independent parameters without loss of generality. That is, $\rho=\rho(\bm{\theta})$ is characterized by $\bm{\theta}$ in Eq.~(\ref{unknown-para}) via Eqs.~(\ref{decom-rho}) and (\ref{alpha-rho}).
It is worth noting that
\begin{center}
    \textit{$\bm{\theta}$ are unknown parameters,}
\end{center}
since they depend on $\rho$ and the only knowledge about $\rho$ is assumed to be the symmetric structures described by $G$ in Theorem 2.

\section{III. Explicit expression of $\overline{X}$}\label{sec-exp-X}
Here we derive an explicit expression of $\overline{X}$ in terms of the parameters $\bm{\theta}$. To do this, we resort to the equality in Theorem 1,
\begin{eqnarray}\label{eq-exp}
    \overline{X}=\overline{Y}.
\end{eqnarray}
Note that $[Y,U_g]=0$ for $g\in G$, which allows us to express $Y$ as
\begin{eqnarray}\label{decom-Y}
    Y\cong\bigoplus_{\alpha=1}^s Y_\alpha\otimes\openone_{d_\alpha},
\end{eqnarray}
with
\begin{eqnarray}\label{Yalp}
    Y_\alpha=a_\alpha\openone_{n_\alpha}+\bm{b}_\alpha\cdot\bm{\lambda}_\alpha.
\end{eqnarray}
Here,
\begin{center}
    \textit{$a_\alpha\in\mathbb{R}$ and $\bm{b}_\alpha\in\mathbb{R}^{n_\alpha^2-1}$ are constants,}
\end{center}
since $Y$ is uniquely determined by the given $X$. Inserting Eqs.~(\ref{decom-rho}), (\ref{alpha-rho}), (\ref{decom-Y}), and (\ref{Yalp}) into Eq.~(\ref{eq-exp}), we obtain
\begin{eqnarray}\label{exp-X}
    \overline{X}=\sum_{\alpha=1}^sq_\alpha (a_\alpha+\bm{b}_\alpha\bm{r}_\alpha^T).
\end{eqnarray}
Equation (\ref{exp-X}) gives an explicit expression of $\overline{X}$ in terms of $\bm{\theta}$.

\section{IV. Proof of Theorem 2} \label{sec2}
Here we prove Theorem 2 by following the methodology of quantum metrology. We divide the proof into three steps.

\textit{First, we set the stage of our analysis.} We treat $\overline{X}$ as the quantity to be estimated.
Resorting to the language of quantum metrology, we can describe a strategy for measuring $\overline{X}$ as first performing a measurement on $\rho^{\otimes M}$ and then inferring the value of $\overline{X}$ from the resultant measurement outcome.
Here, $M$ denotes the number of samples consumed in the measurement strategy in question.
As mentioned in Sec.~I, any measurement can be described by a POVM $\{\Pi_{\bm{x}}\}_{\bm{x}}$, and any inference rule amounts to an estimator $\overline{X}_\textrm{est}(\bm{x})$, which takes the possible measurement outcome $\bm{x}$ of the POVM as its input and outputs an estimate of $\overline{X}$.
Associated with the POVM $\{\Pi_{\bm{x}}\}_{\bm{x}}$ and the (unbiased) estimator $\overline{X}_\textrm{est}(\bm{x})$, the error in measuring $\overline{X}$ can be quantified by the variance $\textrm{Var}[\overline{X}_\textrm{est}]$,
\begin{eqnarray}
    \textrm{Var}[\overline{X}_\textrm{est}]=\sum_{\bm{x}}\tr(\Pi_{\bm{x}}\rho^{\otimes M})\left[\overline{X}_\textrm{est}(\bm{x})-\overline{X}\right]^2.
\end{eqnarray}
The requirement of measuring $\overline{X}$ up to $\epsilon$ amounts to demanding
\begin{eqnarray}\label{requirement}
    \epsilon\geq\textrm{Var}[\overline{X}_\textrm{est}],
\end{eqnarray}
where $\epsilon$ is the prescribed error as mentioned in the main text.

\textit{Second, we introduce a fundamental bound on $M$.} According to the quantum Cram\'{e}r-Rao theorem \cite{1976Helstrom,2011Holevo,1994Braunstein3439} (see Sec.~I), there is
\begin{eqnarray}\label{QCRB}
    \textrm{Var}[\overline{X}_\textrm{est}]\geq\frac{1}{MJ[\overline{X};\rho]},
\end{eqnarray}
where $J[\overline{X};\rho]$ denotes the QFI about $\overline{X}$ given $\rho$.
The meaning of Eq.~(\ref{QCRB}) is that the error produced in any strategy for measuring $\overline{X}$ is fundamentally constrained by $\frac{1}{MJ[\overline{X};\rho]}$, irrespective of the choices of a POVM $\{\Pi_{\bm{x}}\}_{\bm{x}}$ and an estimator $\overline{X}_\textrm{est}(\bm{x})$.
Inserting Eq.~(\ref{requirement}) into Eq.~(\ref{QCRB}), we obtain the following fundamental bound on $M$,
\begin{eqnarray}\label{bound-N}
    M\geq\frac{1}{\epsilon J[\overline{X};\rho]}.
\end{eqnarray}
The explicit expression of $J[\overline{X};\rho]$ is figured out in Sec.~V and is given by
\begin{eqnarray}\label{proof-J}
    &&{J[\overline{X};\rho]}=1\Bigl/\left[\left(\sum_{\alpha=1}^sq_\alpha l_\alpha^2\right)-\left(\sum_{\alpha=1}^sq_\alpha l_\alpha\right)^2\right.\nonumber\\
        &&\left.+\sum_{\alpha=1}^sq_\alpha\bm{b}_\alpha\left(R_\alpha-\bm{r}_\alpha^T\bm{r}_\alpha+\frac{2}{n_\alpha}\openone_{n_\alpha^2-1}\right)\bm{b}_\alpha^T\right],
\end{eqnarray}
where
\begin{eqnarray}
    l_\alpha=a_\alpha+\bm{b}_\alpha\bm{r}_\alpha^T,
\end{eqnarray}
and $R_\alpha$ is a $(n_\alpha^2-1)\times(n_\alpha^2-1)$ symmetric matrix with its $jk$ element defined as
\begin{eqnarray}\label{SM-R}
    R_{\alpha,jk}=\sum_ir_{\alpha,i}g_{\alpha,ijk}.
\end{eqnarray}

\textit{Third, we connect the bound (\ref{bound-N}) to $(\Delta Y)^2$.} Using Eqs.~(\ref{decom-rho}), (\ref{alpha-rho}), (\ref{decom-Y}), and (\ref{Yalp}), we can figure out the explicit expression of $(\Delta Y)^2$ (see Sec.~VI for details), given by
\begin{eqnarray}\label{proof-Y}
    (\Delta Y)^2&=&\sum_{\alpha=1}^sq_\alpha(a_\alpha^2+2a_\alpha\bm{b}_\alpha\bm{r}_\alpha^T+\frac{2}{n_\alpha}\bm{b}_\alpha\bm{b}_\alpha^T+\nonumber\\
    &&\bm{b}_\alpha R_\alpha\bm{b}_\alpha^T)
    -\left(\sum_{\alpha=1}^s q_\alpha l_\alpha\right)^2.
\end{eqnarray}
Rewriting Eq.~(\ref{proof-Y}) by taking into account the equality
\begin{eqnarray}
    a_\alpha^2+2a_\alpha\bm{b}_\alpha\bm{r}_\alpha^T=l_\alpha^2-\bm{b}_\alpha\bm{r}_\alpha^T\bm{r}_\alpha\bm{b}_\alpha^T,
\end{eqnarray}
we obtain
\begin{eqnarray}\label{Y-QFI-new}
    (\Delta Y)^2=\frac{1}{J[\overline{X};\rho]}.
\end{eqnarray}
Inserting Eq.~(\ref{Y-QFI-new}) into Eq.~(\ref{bound-N}) and noting that $M_Y=\left\lceil\frac{(\Delta Y)^2}{\epsilon}\right\rceil$, we have
\begin{eqnarray}
    M\geq M_Y.
\end{eqnarray}
This completes the proof of Theorem 2.

\section{V. Explicit Expression of $J[\overline{X};\rho]$}\label{SM-QFI}
Here we figure out the explicit expression of $J[\overline{X};\rho]$. To this end, we work out the explicit expressions of $H(\bm{\theta})$ and $\partial\overline{X}$  and substituting them into the definition of $J[\overline{X};\rho]$,
\begin{eqnarray}
    J[\overline{X};\rho]=\frac{1}{{\partial\overline{X} \left[H(\bm{\theta})\right]^{-1}\partial\overline{X}^T}}.
\end{eqnarray}
To derive $H(\bm{\theta})$, we figure out the symmetric logarithmic derivatives (SLDs) associated with the unknown parameters $\bm{\theta}=({\bm{q}},\bm{r}_1,\cdots,\bm{r}_s)$.
The SLD associated with $q_\alpha$, denoted by $L_{q_\alpha}$, is the Hermitian operator that satisfies
\begin{eqnarray}\label{eq-SLD-q}
    \pdv{\rho}{q_\alpha}=\rho\circ L_{q_\alpha},
\end{eqnarray}
where $\alpha\in\{1,\cdots,s-1\}$. Noting that the LHS of Eq.~(\ref{eq-SLD-q}) reads
\begin{eqnarray}
    \pdv{\rho}{q_\alpha}\cong\rho_\alpha\otimes\frac{\openone_{d_\alpha}}{d_\alpha}-\rho_s\otimes\frac{\openone_{d_s}}{d_s},
\end{eqnarray}
we have
\begin{eqnarray}\label{SLD-q}
    L_{q_\alpha}\cong\frac{1}{q_\alpha}\openone_{n_\alpha}\otimes\openone_{d_\alpha}-
    \frac{1}{q_s}\openone_{n_s}\otimes\openone_{d_s}.
\end{eqnarray}

The SLD associated with $r_{\alpha,i}$, denoted by $L_{r_{\alpha,i}}$, is the Hermitian operator satisfying
\begin{eqnarray}\label{eq-SLD-r}
    \pdv{\rho}{r_{\alpha,i}}=\rho\circ L_{r_{\alpha,i}}.
\end{eqnarray}
To solve Eq.~(\ref{eq-SLD-r}), we assume the following ansatz for $L_{r_{\alpha,i}}$,
\begin{eqnarray}\label{ansatz}
    L_{r_{\alpha,i}}\cong \left(w_{\alpha,i}\openone_{n_\alpha}+\bm{v}_{\alpha,i}\cdot\bm{\lambda}_\alpha\right)\otimes\openone_{d_\alpha},
\end{eqnarray}
where $w_{\alpha,i}\in\mathbb{R}$ and $\bm{v}_{\alpha,i}\in\mathbb{R}^{n_\alpha^2-1}$ are to be determined.
Besides, it is easy to see that the LHS of Eq.~(\ref{eq-SLD-r}) reads
\begin{eqnarray}\label{lhs-eq-sld-r}
    \pdv{\rho}{r_{\alpha,i}}\cong\frac{1}{2}q_\alpha\lambda_{\alpha,i}\otimes\frac{\openone_{d_\alpha}}{d_\alpha}.
\end{eqnarray}
Substituting Eqs.~(\ref{ansatz}) and (\ref{lhs-eq-sld-r}) into Eq.~(\ref{eq-SLD-r}), we have
\begin{eqnarray}\label{eq-w-v}
    \left(\frac{\openone_{n_\alpha}}{n_\alpha}+\frac{1}{2}\bm{r}_\alpha\cdot\bm{\lambda}_\alpha\right)\circ\left(w_{\alpha,i}\openone_{n_\alpha}+\bm{v}_{\alpha,i}\cdot\bm{\lambda}_\alpha\right)=\frac{1}{2}\lambda_{\alpha,i},\nonumber\\
\end{eqnarray}
which may be viewed as an equation in terms of $w_{\alpha,i}$ and $\bm{v}_{\alpha,i}$. Solving Eq.~(\ref{eq-w-v}) by resorting to the defining properties of $\bm{\lambda}_\alpha$ [specified in Eqs.~(\ref{SM-p1}), (\ref{SM-p2}), and (\ref{SM-p3})], we have
\begin{eqnarray}\label{w}
    w_{\alpha,i}=-\bm{v}_{\alpha,i}\bm{r}_\alpha^T,
\end{eqnarray}
and
\begin{eqnarray}\label{v}
    \bm{v}_{\alpha,i}=\bm{h}_{\alpha,i}\left[R_\alpha-\bm{r}_\alpha^T\bm{r}_\alpha+\frac{2}{n_\alpha}\openone_{n_\alpha^2-1}\right]^{-1}.
\end{eqnarray}
Here, $R_\alpha$ is a $(n_\alpha^2-1)\times(n_\alpha^2-1)$ symmetric matrix with its $jk$ element defined as
\begin{eqnarray}\label{SM-R}
    R_{\alpha,jk}=\sum_ir_{\alpha,i}g_{\alpha,ijk},
\end{eqnarray}
and $\bm{h}_{\alpha,i}$ is a $(n_\alpha^2-1)$-dimensional vector with its $i$-th component identical to one and all others being zero. By the way, we point out that in the case that $[R_\alpha-\bm{r}_\alpha^T\bm{r}_\alpha+\frac{2}{n_\alpha}\openone_{n_\alpha^2-1}]$ is singular, $[R_\alpha-\bm{r}_\alpha^T\bm{r}_\alpha+\frac{2}{n_\alpha}\openone_{n_\alpha^2-1}]^{-1}$is understood as the Moore–Penrose inverse
of $[R_\alpha-\bm{r}_\alpha^T\bm{r}_\alpha+\frac{2}{n_\alpha}\openone_{n_\alpha^2-1}]$.

It is easy to see that for three matrices $A$, $B$, and $C$, there are
\begin{eqnarray}\label{eq-jordan}
    A\circ B= B\circ A,~~~~\tr[A(B\circ C)]=\tr[(A\circ B) C].
\end{eqnarray}
Using these two equalities and Eqs.~(\ref{SLD-q}), (\ref{eq-SLD-r}), and (\ref{lhs-eq-sld-r}), we have
\begin{eqnarray}\label{vanishing1}
    &&\tr\left[\rho\left(L_{q_\alpha}\circ L_{r_{\beta,i}}\right)\right]\nonumber\\
    &=&
    \tr\left[\left(\rho\circ L_{r_{\beta,i}}\right)L_{q_\alpha}\right]\nonumber\\
    &=&\tr(\pdv{\rho}{r_{\beta,i}}L_{q_\alpha})\nonumber\\
    &\cong&\tr\left[\frac{1}{2}q_\beta\lambda_{\beta,i}\otimes\frac{\openone_{d_\beta}}{d_\beta}\left(\frac{1}{q_\alpha}\openone_{n_\alpha}\otimes\openone_{d_\alpha}-
        \frac{1}{q_s}\openone_{n_s}\otimes\openone_{d_s}\right)\right]\nonumber\\
    &=&0.
\end{eqnarray}
Here, $\beta$ is a subscript and should be distinguished from the notation used in Sec.~I. Using Eq.~(\ref{ansatz}), we easily have
\begin{eqnarray}\label{vanishing2}
    \tr\left[\rho\left(L_{r_{\alpha,i}}\circ L_{r_{\beta,j}}\right)\right]=0,
\end{eqnarray}
for $\alpha\neq\beta$.

From Eqs.~(\ref{vanishing1}) and (\ref{vanishing2}), we deduce that $H(\bm{\theta})$ can be expressed in a block-diagonal form
\begin{eqnarray}\label{QFI-matrix-2}
    H(\bm{\theta})=H({\bm{q}})\bigoplus\left[\bigoplus_{\alpha=1}^s H(\bm{r}_\alpha)\right],
\end{eqnarray}
where $H({\bm{q}})$ is a $(s-1)\times(s-1)$ symmetric matrix with its $\alpha\beta$ element defined as
\begin{eqnarray}\label{QFI-q}
    \left[H({\bm{q}})\right]_{\alpha\beta}=\tr[\rho (L_{q_\alpha}\circ L_{q_\beta})],
\end{eqnarray}
and $H(\bm{r}_\alpha)$ is a $(n_\alpha^2-1)\times(n_\alpha^2-1)$ symmetric matrix with its $ij$ element defined as
\begin{eqnarray}\label{QFI-r}
    \left[H(\bm{r}_\alpha)\right]_{ij}=\tr[\rho \left(L_{r_{\alpha,i}}\circ L_{r_{\alpha,j}}\right)].
\end{eqnarray}
Inserting Eq.~(\ref{SLD-q}) into Eq.~(\ref{QFI-q}), we have, after simple algebra,
\begin{eqnarray}
    \left[H({\bm{q}})\right]_{\alpha\beta}=\delta_{\alpha\beta}\frac{1}{q_\alpha}+\frac{1}{q_s},
\end{eqnarray}
that is,
\begin{eqnarray}\label{H-q-ex}
    H({\bm{q}})=\textrm{diag}(\frac{1}{q_1},\cdots,\frac{1}{q_{s-1}})+\frac{1}{q_s}\bm{e}^T\bm{e},
\end{eqnarray}
where $\bm{e}$ is a $(s-1)$-dimensional vector with all components identical to one.
Substituting Eqs.~(\ref{ansatz}), (\ref{lhs-eq-sld-r}), (\ref{w}), and (\ref{v}) into Eq.~(\ref{QFI-r}) gives
\begin{eqnarray}
    &&\left[H(\bm{r}_\alpha)\right]_{ij}\nonumber\\
    &=&\tr[\rho \left(L_{r_{\alpha,i}}\circ L_{r_{\alpha,j}}\right)]\nonumber\\
    &=&\tr[\left(\rho\circ L_{r_{\alpha,i}}\right)L_{r_{\alpha,j}}]\nonumber\\
    &=&\tr[\pdv{\rho}{r_{\alpha,i}}L_{r_{\alpha,j}}]\nonumber\\
    &=&\tr[\left(\frac{1}{2}q_\alpha\lambda_{\alpha,i}\otimes\frac{\openone_{d_\alpha}}{d_\alpha}\right)\left(w_{\alpha,j}\openone_{n_\alpha}+\bm{v}_{\alpha,j}\cdot\bm{\lambda}_\alpha\right)\otimes\openone_{d_\alpha}]\nonumber\\
    &=&q_\alpha\bm{h}_{\alpha,i}\bm{v}_{\alpha,j}^T\nonumber\\
    &=&q_\alpha\bm{h}_{\alpha,i}\left[R_\alpha-\bm{r}_\alpha^T\bm{r}_\alpha+\frac{2}{n_\alpha}\openone_{n_\alpha^2-1}\right]^{-1}\bm{h}_{\alpha,j}^T.
\end{eqnarray}
That is,
\begin{eqnarray}\label{H-r-ex}
    H(\bm{r}_\alpha)=q_\alpha \left[R_\alpha-\bm{r}_\alpha^T\bm{r}_\alpha+\frac{2}{n_\alpha}\openone_{n_\alpha^2-1}\right]^{-1}.
\end{eqnarray}
Now, $H(\bm{\theta})$ is given by Eqs.~(\ref{QFI-matrix-2}) with Eqs.~(\ref{H-q-ex}) and (\ref{H-r-ex}).

On the other hand, we have shown that $\overline{X}$ can be expressed as
\begin{eqnarray}
    \overline{X}=\sum_{\alpha=1}^sq_\alpha l_\alpha,
\end{eqnarray}
with
\begin{eqnarray}
    l_\alpha=a_\alpha+\bm{b}_\alpha\bm{r}_\alpha^T.
\end{eqnarray}
Differentiating  $\overline{X}$ with respect to $q_\alpha$ where $\alpha\in\{1,\cdots,s-1\}$, we have
\begin{eqnarray}
    \frac{\partial\overline{X}}{\partial q_\alpha}=l_\alpha-l_s,
\end{eqnarray}
where we have used the fact that
\begin{eqnarray}
    q_s=1-\sum_{\alpha=1}^{s-1}q_\alpha.
\end{eqnarray}
Then, differentiating $\overline{X}$ with respect to $r_{\alpha,i}$, we have
\begin{eqnarray}
    \frac{\partial\overline{X}}{\partial r_{\alpha,i}}=q_\alpha b_{\alpha,i}.
\end{eqnarray}
Therefore,
\begin{eqnarray}\label{partial-beta}
    \partial\overline{X}=\left(\bm{l}-l_s\bm{e},q_1\bm{b}_1,\cdots,q_s\bm{b}_s\right),
\end{eqnarray}
where
\begin{eqnarray}
    \bm{l}=(l_1,\cdots,l_{s-1}).
\end{eqnarray}

Using Eqs.~(\ref{QFI-matrix-2}) and (\ref{partial-beta}), we obtain
\begin{eqnarray}\label{SM-left-new}
    &&{J[\overline{X};\rho]}=1\Bigl/\left[\left(\sum_{\alpha=1}^sq_\alpha l_\alpha^2\right)-\left(\sum_{\alpha=1}^sq_\alpha l_\alpha\right)^2\right.\nonumber\\
        &&\left.+\sum_{\alpha=1}^sq_\alpha\bm{b}_\alpha\left(R_\alpha-\bm{r}_\alpha^T\bm{r}_\alpha+\frac{2}{n_\alpha}\openone_{n_\alpha^2-1}\right)\bm{b}_\alpha^T\right],
\end{eqnarray}
where we have used the equality
\begin{eqnarray}
    [H({\bm{q}})]^{-1}=\textrm{diag}(q_1,\cdots,q_{s-1})-{\bm{q}}^T{\bm{q}}.
\end{eqnarray}

\section{VI. Explicit expression of $(\Delta Y)^2$}\label{sec4}

Here we figure out the explicit expression of $(\Delta Y)^2$. We deduce from Eqs.~(\ref{decom-rho}) and (\ref{decom-Y}) that
\begin{eqnarray}
    \overline{Y^2}=\sum_{\alpha=1}^sq_\alpha\tr(\rho_\alpha Y_\alpha^2).
\end{eqnarray}
Then, using Eqs.~(\ref{alpha-rho}) and (\ref{Yalp}) and noting that
\begin{eqnarray}
    \tr(\bm{b}_\alpha\cdot\bm{\lambda}_\alpha\bm{r}_\alpha\cdot\bm{\lambda}_\alpha)=2\bm{b}_\alpha\bm{r}_\alpha^T,
\end{eqnarray}
\begin{eqnarray}
    \tr(\bm{b}_\alpha\cdot\bm{\lambda}_\alpha\bm{b}_\alpha\cdot\bm{\lambda}_\alpha)=2\bm{b}_\alpha\bm{b}_\alpha^T,
\end{eqnarray}
\begin{eqnarray}
    \tr(\bm{b}_\alpha\cdot\bm{\lambda}_\alpha\bm{r}_\alpha\cdot\bm{\lambda}_\alpha\bm{b}_\alpha\cdot\bm{\lambda}_\alpha)=2\bm{b}_\alpha R_\alpha\bm{b}_\alpha^T,
\end{eqnarray}
we have
\begin{eqnarray}
    \overline{Y^2}=\sum_{\alpha=1}^sq_\alpha\left(a_\alpha^2+2a_\alpha\bm{b}_\alpha\bm{r}_\alpha^T+\frac{2}{n_\alpha}\bm{b}_\alpha\bm{b}_\alpha^T+\bm{b}_\alpha R_\alpha\bm{b}_\alpha^T\right).\nonumber
\end{eqnarray}
Using this equality and noting that $\overline{Y}=\sum_{\alpha=1}^s q_\alpha l_\alpha$, we obtain
\begin{eqnarray}\label{SM-right}
    (\Delta Y)^2&=&\sum_{\alpha=1}^sq_\alpha(a_\alpha^2+2a_\alpha\bm{b}_\alpha\bm{r}_\alpha^T+\frac{2}{n_\alpha}\bm{b}_\alpha\bm{b}_\alpha^T+\nonumber\\
    &&\bm{b}_\alpha R_\alpha\bm{b}_\alpha^T)
    -\left(\sum_{\alpha=1}^s q_\alpha l_\alpha\right)^2.
\end{eqnarray}
As shown in Sec.~IV, Eq.~(\ref{SM-right}) can be rewritten by taking into account the equality
\begin{eqnarray}
    a_\alpha^2+2a_\alpha\bm{b}_\alpha\bm{r}_\alpha^T=l_\alpha^2-\bm{b}_\alpha\bm{r}_\alpha^T\bm{r}_\alpha\bm{b}_\alpha^T,
\end{eqnarray}
which gives
\begin{eqnarray}\label{Y-QFI}
    (\Delta Y)^2=\frac{1}{J[\overline{X};\rho]}.
\end{eqnarray}

%\part{\large{Calculation details on illustrative applications}}

\section{VII. Details on application 1}\label{sec6}

Here we prove the two inequality used in the main text,
\begin{eqnarray}\label{app1-ratio}
    \frac{(\Delta X)^2}{(\Delta Y)^2}\geq 2^n-1.
\end{eqnarray}
We resort to the product representation of the Fourier basis \cite{2010Nielsen},
\begin{eqnarray}\label{SM-PR}
    \ket{f_j}=\frac{\left(\ket{0}+e^{2\pi i0.j_n}\ket{1}\right)\cdots\left(\ket{0}+e^{2\pi i0.j_1\cdots j_n}\ket{1}\right)}{2^{n/2}},\nonumber\\
\end{eqnarray}
where $0.j_lj_{l+1}\cdots j_m$ denotes the binary fraction $j_l/2+j_{l+1}/4+\cdots+j_m/2^{m-l+1}$. Using Eq.~(\ref{SM-PR}) to calculate $X_{jj}=\bra{f_j}X\ket{f_j}$ gives
\begin{eqnarray}
    X_{jj}=\frac{1}{\sqrt{2^n}}\cos(2\pi 0.j_n)\cdots\cos(2\pi 0.j_1j_2\cdots j_n).
\end{eqnarray}
Noting that $\abs{X_{jj}}\leq \frac{1}{\sqrt{2^n}}$ and $\rho=\sum_jp_j\ket{f_j}\bra{f_j}$, we have
\begin{eqnarray}\label{SM-App1-X-expval}
    \abs{\overline{X}}=\abs{\sum_jp_jX_{jj}}\leq\sum_jp_j\abs{X_{jj}}\leq \frac{1}{\sqrt{2^n}}\sum_jp_j=\frac{1}{\sqrt{2^n}}.\nonumber
\end{eqnarray}
Besides, note that $X^2=\openone\otimes\cdots\otimes\openone$ and therefore
\begin{eqnarray}\label{SM-App1-X2-expval}
    \overline{X^2}=1,
\end{eqnarray}
which further leads to
\begin{eqnarray}\label{SM-App1-Eq1}
    1-\frac{1}{2^n}\leq (\Delta X)^2\leq 1.
\end{eqnarray}
On the other hand, noting that $Y=\sum_jX_{jj}\ket{f_j}\bra{f_j}$, we have
\begin{eqnarray}\label{SM-App1-Y-expval}
    \abs{\overline{Y}}=\abs{\overline{X}}\leq \frac{1}{\sqrt{2^n}},
\end{eqnarray}
and
\begin{eqnarray}\label{SM-App1-Y2-expval}
    \overline{Y^2}=\sum_jp_jX_{jj}^2\leq \frac{1}{2^n}.
\end{eqnarray}
Therefore,
\begin{eqnarray}\label{SM-App1-Eq2}
    0\leq (\Delta Y)^2\leq \frac{1}{2^n}.
\end{eqnarray}
Then Eq.~(\ref{app1-ratio}) follows from Eqs.~(\ref{SM-App1-Eq1}) and (\ref{SM-App1-Eq2}).

\section{VIII. Details on application 2}\label{sec7}
\subsection{Calculating $(\Delta X_{\bm{kl}})^2$ and $(\Delta Y_{\bm{kl}})^2$}

Here we figure out the expressions of $(\Delta X_{\bm{kl}})^2$ and $(\Delta Y_{\bm{kl}})^2$. Note that
\begin{eqnarray}\label{ap2:tensor-GHZ}
    &&\sigma_x^{k_1}\sigma_z^{l_1}\otimes\cdots\otimes\sigma_x^{k_n}\sigma_z^{l_n}
    \ket{GHZ}=\nonumber\\
    &&\frac{1}{\sqrt{2}}\left[\ket{k_1,\cdots,k_n}+(-1)^{\abs{\bm{l}}}\ket{1-k_1,\cdots,1-k_n}\right],\nonumber\\
\end{eqnarray}
with the $n$-qubit GHZ state $\ket{GHZ}=\left(\ket{0\cdots 0}+\ket{1\cdots 1}\right)/\sqrt{2}$ and $\abs{\bm{l}}=\sum_{i=1}^nl_i$.
Using Eq.~(\ref{ap2:tensor-GHZ}) and
\begin{eqnarray}
    X_{\bm{kl}}=\sigma_x^{k_1}\sigma_z^{l_1}\otimes\cdots\otimes\sigma_x^{k_n}\sigma_z^{l_n}(i)^{\bm{k}\cdot\bm{l}},
\end{eqnarray}
we have
\begin{eqnarray}
    \bra{GHZ}X_{\bm{kl}}\ket{GHZ}=(i)^{\bm{k}\cdot\bm{l}}\frac{1+(-1)^{\abs{\bm{l}}}}{2}(\delta_{\bm{k},\bm{0}}+\delta_{\bm{k},\bm{1}}),\nonumber\\
\end{eqnarray}
where $\delta_{\bm{k},\bm{0}}:=\delta_{k_1,0}\cdots\delta_{k_n,0}$ and $\delta_{\bm{k},\bm{1}}:=\delta_{k_1,1}\cdots\delta_{k_n,1}$. Hence,
\begin{eqnarray}\label{ap2:exp-X}
    &&\bra{GHZ}X_{\bm{kl}}\ket{GHZ}=\nonumber\\
    &&\begin{cases}
        1                  & \bm{k}=\bm{0} ~~\textrm{and}~~ \abs{\bm{l}}\in \textrm{even numbers}, \\
        (i)^{\abs{\bm{l}}} & \bm{k}=\bm{1} ~~\textrm{and}~~ \abs{\bm{l}}\in \textrm{even numbers}, \\
        0                  & \textrm{otherwise}.
    \end{cases}
\end{eqnarray}
It follows from Eq.~(\ref{ap2:exp-X}) and $X_{\bm{kl}}^2=\openone_{2^n}$ that
\begin{eqnarray}
    (\Delta X_{\bm{kl}})^2=
    \begin{cases}
        0 & \bm{k}=\bm{0},\bm{1} ~~\textrm{and}~~ \abs{\bm{l}}\in \textrm{even numbers}, \\
        1 & \textrm{otherwise}.
    \end{cases}\nonumber\\
\end{eqnarray}
To calculate $(\Delta Y_{\bm{kl}})^2$, we resort to the defining properties of $P_s$,
\begin{eqnarray}
    P_s\ket{\psi_1}\otimes\cdots\otimes\ket{\psi_n}=
    \ket{\psi_{s(1)}}\otimes\cdots\otimes\ket{\psi_{s(n)}}
\end{eqnarray}
and
\begin{eqnarray}
    P_s A_1\otimes\cdots\otimes A_n P_s^\dagger=
    A_{s(1)}\otimes\cdots\otimes A_{s(n)},
\end{eqnarray}
which, in conjunction with the rearrangement theorem, lead to
\begin{eqnarray}\label{ap2:properties}
    P_s\ket{GHZ}=\ket{GHZ}, ~~P_s Y_{\bm{kl}}=Y_{\bm{kl}}P_s.
\end{eqnarray}
Using Eq.~(\ref{ap2:properties}) and
\begin{eqnarray}
    &&Y_{\bm{kl}}=\frac{1}{n!}\sum_{s\in S_n}\sigma_x^{k_{s(1)}}\sigma_z^{l_{s(1)}}\otimes\cdots\otimes\sigma_x^{k_{s(n)}}\sigma_z^{l_{s(n)}} (i)^{\bm{k}\cdot\bm{l}},\nonumber\\
\end{eqnarray}
we have
\begin{widetext}
    \begin{eqnarray}\label{ap2:Y-square}
        \bra{GHZ}Y_{\bm{kl}}^2\ket{GHZ}
        &=&\frac{1}{n!}\sum_{s\in S_n}\bra{GHZ}\sigma_x^{k_{s(1)}}\sigma_z^{l_{s(1)}}\otimes\cdots\otimes\sigma_x^{k_{s(n)}}\sigma_z^{l_{s(n)}} (i)^{\bm{k}\cdot\bm{l}}Y_{\bm{kl}}\ket{GHZ}\nonumber\\
        &=&
        \frac{1}{n!}\sum_{s\in S_n}\bra{GHZ}P_s^\dagger\sigma_x^{k_{s(1)}}\sigma_z^{l_{s(1)}}\otimes\cdots\otimes\sigma_x^{k_{s(n)}}\sigma_z^{l_{s(n)}} (i)^{\bm{k}\cdot\bm{l}}Y_{\bm{kl}}P_s\ket{GHZ}\nonumber\\
        &=&
        \frac{1}{n!}\sum_{s\in S_n}\bra{GHZ}P_s^\dagger\sigma_x^{k_{s(1)}}\sigma_z^{l_{s(1)}}\otimes\cdots\otimes\sigma_x^{k_{s(n)}}\sigma_z^{l_{s(n)}} (i)^{\bm{k}\cdot\bm{l}}P_s Y_{\bm{kl}}\ket{GHZ}\nonumber\\
        &=&
        \frac{1}{n!}\sum_{s\in S_n}\bra{GHZ}\sigma_x^{k_{1}}\sigma_z^{l_{1}}\otimes\cdots\otimes\sigma_x^{k_{n}}\sigma_z^{l_{n}} (i)^{\bm{k}\cdot\bm{l}}Y_{\bm{kl}}\ket{GHZ}\nonumber\\
        &=&
        \bra{GHZ}\sigma_x^{k_{1}}\sigma_z^{l_{1}}\otimes\cdots\otimes\sigma_x^{k_{n}}\sigma_z^{l_{n}} (i)^{\bm{k}\cdot\bm{l}}Y_{\bm{kl}}\ket{GHZ}\nonumber\\
        &=&\frac{1}{n!}\sum_{s\in S_n}
        \bra{GHZ}\sigma_x^{k_{1}}\sigma_z^{l_{1}}\sigma_x^{k_{s(1)}}\sigma_z^{l_{s(1)}}\otimes\cdots\otimes\sigma_x^{k_{n}}\sigma_z^{l_{n}}\sigma_x^{k_{s(n)}}\sigma_z^{l_{s(n)}}(-1)^{\bm{k}\cdot\bm{l}}\ket{GHZ}\nonumber\\
        &=&\frac{1}{n!}\sum_{s\in S_n}
        \bra{GHZ}\sigma_x^{k_{1}+k_{s(1)}}\sigma_z^{l_{1}+l_{s(1)}}\otimes\cdots\otimes\sigma_x^{k_{n}+k_{s(n)}}\sigma_z^{l_{n}+l_{s(n)}}(-1)^{(\bm{k}+\bm{k}_s)\cdot\bm{l}}\ket{GHZ}\nonumber\\
        &=&\frac{1}{n!}\sum_{s\in S_n}(-1)^{(\bm{k}+\bm{k}_s)\cdot\bm{l}}\frac{1+(-1)^{\abs{\bm{l}+\bm{l}_s}}}{2}
        (\delta_{\bm{k}\oplus\bm{k}_s,\bm{0}}+\delta_{\bm{k}\oplus\bm{k}_s,\bm{1}}),
    \end{eqnarray}
\end{widetext}
where $\bm{k}_s=(k_{s(1)},\cdots,k_{s(n)})$, $\bm{l}_s=(l_{s(1)},\cdots,l_{s(n)})$, and $\oplus$ denotes addition modulo $2$, that is, $\bm{k}\oplus\bm{k}_s=(k_1+k_{s(1)}, \cdots, k_n+k_{s(n)})$ ({mod} $2$). Noting that $\abs{\bm{l}+\bm{l}_s}=\sum_{i=1}^nl_i+l_{s(i)}=2\abs{\bm{l}}$, we have $[1+(-1)^{\abs{\bm{l}+\bm{l}_s}}]/{2}=1$. Moreover, assuming that $n$ is odd, we have $\delta_{\bm{k}\oplus\bm{k}_s,\bm{1}}=0$. So, the only nonzero terms in the summation of Eq.~(\ref{ap2:Y-square}) are those such that $\bm{k}=\bm{k}_s$. The number of such terms is $\abs{\bm{k}}!(n-\abs{\bm{k}})!$. Besides, $(-1)^{(\bm{k}+\bm{k}_s)\cdot\bm{l}}=1$ when $\bm{k}=\bm{k}_s$. We have
\begin{eqnarray}\label{ap2:Y-sq}
    \bra{GHZ}Y_{\bm{kl}}^2\ket{GHZ}=1/\binom{n}{\abs{\bm{k}}}.
\end{eqnarray}
Noting that $\bra{GHZ}Y_{\bm{kl}}\ket{GHZ}=\bra{GHZ}X_{\bm{kl}}\ket{GHZ}$ and using Eqs.~(\ref{ap2:exp-X}) and (\ref{ap2:Y-sq}), we arrive at the expression of $(\Delta Y_{\bm{kl}})^2$:
\begin{eqnarray}
    (\Delta Y_{\bm{kl}})^2=
    \begin{cases}
        0                         & \bm{k}=\bm{0}, \bm{1} ~~\textrm{and}~~ \abs{\bm{l}}\in \textrm{even numbers}, \\
        1/\binom{n}{\abs{\bm{k}}} & \textrm{otherwise}.
    \end{cases}\nonumber\\
\end{eqnarray}

\subsection{Lower bound on $(\Delta X)^2/(\Delta Y)^2$ for typical Pauli measurements}

Here we figure out the lower bound on $(\Delta X)^2/(\Delta Y)^2$ for typical Pauli measurements. Using the Stirling's approximation
\begin{eqnarray}
    n!\approx\sqrt{2\pi n}\left(\frac{n}{e}\right)^n
\end{eqnarray}
and noting that
\begin{eqnarray}
    (1-\delta)\frac{n}{2}
    <\abs{\bm{k}}<(1+\delta)\frac{n}{2},
\end{eqnarray}
we have
\begin{widetext}
    \begin{eqnarray}
        {(\Delta X)^2}/{(\Delta Y)^2}
        &=&\binom{n}{\abs{\bm{k}}}\nonumber\\
        &\geq&\binom{n}{(1-\delta)\frac{n}{2}}\nonumber\\
        &=&\frac{n!}{\left[(1-\delta)\frac{n}{2}\right]!\left[(1+\delta)\frac{n}{2}\right]!}\nonumber\\
        &\approx&\frac{\sqrt{2\pi n}\left(\frac{n}{e}\right)^n}{\sqrt{2\pi (1-\delta)\frac{n}{2}}\left(\frac{(1-\delta)\frac{n}{2}}{e}\right)^{(1-\delta)\frac{n}{2}}
            \sqrt{2\pi (1+\delta)\frac{n}{2}}\left(\frac{(1+\delta)\frac{n}{2}}{e}\right)^{(1+\delta)\frac{n}{2}}}\nonumber\\
        &=&\sqrt{\frac{2}{n\pi(1-\delta^2)}}\left[\frac{4}{(1-\delta)^{1-
                        \delta}(1+\delta)^{1+
                        \delta}}\right]^{\frac{n}{2}}.
    \end{eqnarray}
\end{widetext}
Here we have assumed for simplicity that $(1-\delta)\frac{n}{2}$ is an integer. The above proof can be made more rigorous by replacing the  Stirling's approximation with the inequalities $\sqrt{2\pi n}\left(\frac{n}{e}\right)^ne^{\frac{1}{12n+1}}\leq n!\leq \sqrt{2\pi n}\left(\frac{n}{e}\right)^ne^{\frac{1}{12n}}$ which hold for $n\geq 1$.

\subsection{Lower bound on the number of typical Pauli measurements}
Here we find a lower bound on the number of typical Pauli measurements. To do this, we interpret all $k_i$'s as independent and identically distributed binary random variables taking on two values $0$ and $1$ with equal probabilities $1/2$. Then, all the bit strings $\bm{k}=(k_1,\cdots,k_n)$ are with equal probabilities $1/2^n$. Using the Chebyshev's inequality \cite{Mitzenmacher2005} and noting that $\textrm{Var}[\abs{\bm{k}}]=\textrm{Var}[k_1+\cdots+k_n]=n/4$, we have that the number of the bit strings with $(1-\delta)\frac{n}{2}
    <\abs{\bm{k}}<(1+\delta)\frac{n}{2}$, denoted as $N$, is
\begin{eqnarray}\label{SM-App2-N}
    N&=&2^n\textrm{Pr}\left((1-\delta)\frac{n}{2}
    <\abs{\bm{k}}<(1+\delta)\frac{n}{2}\right)\nonumber\\
    &=&2^n\left[1-\textrm{Pr}\left(\abs{\abs{\bm{k}}-\frac{n}{2}}\geq \frac{n\delta}{2}\right)\right]\nonumber\\
    &\geq&2^n\left[1-\frac{\textrm{Var}[\abs{\bm{k}}]}{\left(\frac{n\delta}{2}\right)^2}\right]\nonumber\\
    &=&2^n\left(1-\frac{1}{n\delta^2}\right),
\end{eqnarray}
where $\textrm{Pr}\left((1-\delta)\frac{n}{2}
    <\abs{\bm{k}}<(1+\delta)\frac{n}{2}\right)$ stands for the probability that a bit string $\bm{k}$ is with $(1-\delta)\frac{n}{2}
    <\abs{\bm{k}}<(1+\delta)\frac{n}{2}$, and similarly for the notation $\textrm{Pr}\left(\abs{\abs{\bm{k}}-\frac{n}{2}}\geq \frac{n\delta}{2}\right)$. The number of typical Pauli measurements is
\begin{eqnarray}
    2^n N\geq 4^n\left(1-\frac{1}{n\delta^2}\right),
\end{eqnarray}
which follows from the fact that there are $2^n$ different $\bm{l}$ for each $\bm{k}$.

\subsection{Discussion on the implementation of $V$}
Here we explain why the quantum circuit $V$ can be efficiently realized as a quantum circuit. According to the Schur–Weyl duality \cite{Weyl1950}, the Hilbert space $(\mathbb{C}^2)^{\otimes n}$ of $n$ qubits can be decomposed into a direct sum of tensor products,
\begin{eqnarray}\label{SM-decom-H}
    (\mathbb{C}^2)^{\otimes n}\cong\bigoplus_{j=j_\textrm{min}}^{j_\textrm{max}}\mathbb{C}^{2j+1}\otimes \mathbb{C}^{d_{n,j}},
\end{eqnarray}
where $\mathbb{C}^{2j+1}$ and $\mathbb{C}^{d_{n,j}}$ are irreducible representations of $U(2)$ and $S_n$, respectively. $j$ labels the eigen-subspace of the total squared angular
momentum with eigenvalue $j(j+1)$, where $j=j_\textrm{min}$, $j_\textrm{min}+1$, $\cdots$, $j_\textrm{max}$ with $j_\textrm{max}=n/2$ and  $j_\textrm{min}=0$, $1/2$ for even and odd $n$, respectively.
The decomposition (\ref{SM-decom-H}) corresponds to a basis for $(\mathbb{C}^2)^{\otimes n}$, $\ket{\Psi_{j,q,p}}$, known as the Schur basis \cite{Bacon2006}, where $j$ labels the subspaces $\mathbb{C}^{2j+1}\otimes \mathbb{C}^{d_{n,j}}$, and $q$ and $p$ label bases for $\mathbb{C}^{2j+1}$ and $\mathbb{C}^{d_{n,j}}$, respectively. The Schur transform, denoted as $U_\textrm{Sch}$ hereafter, is defined as the unitary operator mapping the Schur basis into the computational basis,
\begin{eqnarray}\label{SM-Schur}
    U_\textrm{Sch}\ket{\Psi_{j,q,p}}=\ket{j,q,p}.
\end{eqnarray}
Here, $\ket{j,q,p}$ stands for the computational basis, with $j,q,p$ expressed as bit strings.
A central result of Ref.~\cite{Bacon2006} is that $U_\textrm{Sch}$ can be realized as a quantum circuit of polynomial size.

Without loss of generality, we can express $X$ in the form
\begin{eqnarray}\label{SM-App2-X}
    X=\sum_{j,j^\prime}\sum_{q,q^\prime}\sum_{p,p^\prime}X_{jqp;j^\prime q^\prime p^\prime}\ket{\Psi_{j,q,p}}\bra{\Psi_{j^\prime,q^\prime,p^\prime}},
\end{eqnarray}
where
\begin{eqnarray}
    X_{jqp;j^\prime q^\prime p^\prime}=\bra{\Psi_{j,q,p}}X\ket{\Psi_{j^\prime,q^\prime,p^\prime}}.
\end{eqnarray}
Moreover, noting that $\rho$ and $Y$ respect the permutation symmetries, we can express $\rho$ and $Y$ in the form
\begin{eqnarray}
    \rho=\sum_j\sum_{q,q^\prime}\sum_pp_j(\rho_j)_{qq^\prime}\ket{\Psi_{j,q,p}}\bra{\Psi_{j,q^\prime,p}}
\end{eqnarray}
and
\begin{eqnarray}\label{SM-App2-Y}
    Y=\sum_j\sum_{q,q^\prime}\sum_p(Y_j)_{qq^\prime}\ket{\Psi_{j,q,p}}\bra{\Psi_{j,q^\prime,p}},
\end{eqnarray}
where $p_j\geq 0$ satisfies $\sum_jp_j=1$, $\rho_j$ is a density matrix, and $Y_j$ is a Hermitian matrix. To find the connection between $X$ and $Y$, we make use of the equality $\overline{X}=\overline{Y}$, which leads to
\begin{eqnarray}
    ({Y}_{j})_{qq^\prime}=\frac{1}{d_{n,j}}\sum_pX_{jqp;j q^\prime p}.
\end{eqnarray}
It follows from Eqs.~(\ref{SM-Schur}) and (\ref{SM-App2-Y}) that
\begin{eqnarray}
    &&U_\textrm{Sch}YU_\textrm{Sch}^\dagger=\sum_{j}\sum_{q,q^\prime}\sum_p(Y_{j})_{qq^\prime}\ket{{j,q,p}}\bra{{j,q^\prime,p}}\nonumber\\
    &&=\sum_{j}\ket{j}\bra{j}\otimes\left(\sum_{q,q^\prime}(Y_{j})_{qq^\prime}\ket{q}\bra{q^\prime}\right)\otimes\openone_{d_{n,j}}.\nonumber\\
\end{eqnarray}
Let the eigendecomposition  of $Y_j$ be
\begin{eqnarray}
    Y_j=U_jD_jU_j^\dagger,
\end{eqnarray}
where $U_j$ denotes a unitary matrix and $D_j$ is a diagonal matrix with real entries. We then define $V$ to be of the form
\begin{eqnarray}
    V=V_{j_\textrm{min}}V_{j_\textrm{min}+1}\cdots V_{j_\textrm{max}} U_\textrm{Sch},
\end{eqnarray}
where
\begin{eqnarray}
    V_j=\ket{j}\bra{j}\otimes U_j^\dagger+\sum_{j^\prime\neq j}\ket{j^\prime}\bra{j^\prime}\otimes\openone_{2j+1}
\end{eqnarray}
is a controlled gate. It is easy to see that the $V$ thus defined transforms the eigenbasis of $Y$ into the computational basis $\ket{j,q,p}$. As each $V_j$ is a unitary matrix of polynomial size, $V$ can be realized as a quantum circuit of polynomial size.

%\bibliography{refs.bib}
%apsrev4-2.bst 2019-01-14 (MD) hand-edited version of apsrev4-1.bst
%Control: key (0)
%Control: author (8) initials jnrlst
%Control: editor formatted (1) identically to author
%Control: production of article title (0) allowed
%Control: page (0) single
%Control: year (1) truncated
%Control: production of eprint (0) enabled
%

\end{document}